\documentclass[12pt]{article} 
\usepackage[utf8x]{inputenc} 
\usepackage[T1]{fontenc} 
\usepackage{amsmath,amssymb,bbm,bm,commath,mathtools,slashed,wasysym,xfrac} 
\usepackage{sectsty} 
\usepackage{graphicx} 
\graphicspath{{figures/}}
\usepackage{color} 
\usepackage{cite} 

\usepackage{geometry} 
\usepackage{fancyhdr} 
\usepackage{setspace} 
\usepackage[nottoc,notlof,notlot]{tocbibind} 
\usepackage[titles]{tocloft} 

\usepackage[unicode]{hyperref} 
\hypersetup{bookmarksnumbered=true, bookmarksopen=true, bookmarksopenlevel=2, breaklinks=true, citecolor=blue, colorlinks=true, linkcolor=red, linktoc=page, pdfborder={0 0 0}, pdfstartview=FitH, pdfauthor={DJain}, pdfsubject={Twisted Indices of CS Quiver Theories}, pdftitle={Twisted Indices of more 3d Quivers}, plainpages=false, unicode=true, urlcolor=cyan}

\DeclareUnicodeCharacter{"0393}{\Gamma}
\DeclareUnicodeCharacter{"0394}{\Delta}
\DeclareUnicodeCharacter{"0398}{\Theta}
\DeclareUnicodeCharacter{"039B}{\Lambda}
\DeclareUnicodeCharacter{"039E}{\Xi}
\DeclareUnicodeCharacter{"03A0}{\Pi}
\DeclareUnicodeCharacter{"03A3}{\Sigma}
\DeclareUnicodeCharacter{"03A5}{\Upsilon}
\DeclareUnicodeCharacter{"03A6}{\Phi}
\DeclareUnicodeCharacter{"03A8}{\Psi}
\DeclareUnicodeCharacter{"03A9}{\Omega}
\DeclareUnicodeCharacter{"03B1}{\alpha}
\DeclareUnicodeCharacter{"03B2}{\beta}
\DeclareUnicodeCharacter{"03B3}{\gamma}
\DeclareUnicodeCharacter{"03B4}{\delta}
\DeclareUnicodeCharacter{"03B5}{\epsilon}
\DeclareUnicodeCharacter{"03B6}{\zeta}
\DeclareUnicodeCharacter{"03B7}{\eta}
\DeclareUnicodeCharacter{"03B8}{\theta}
\DeclareUnicodeCharacter{"03D1}{\vartheta}
\DeclareUnicodeCharacter{"03B9}{\iota}
\DeclareUnicodeCharacter{"03BA}{\kappa}
\DeclareUnicodeCharacter{"03BB}{\lambda}
\DeclareUnicodeCharacter{"03BC}{\mu}
\DeclareUnicodeCharacter{"03BD}{\nu}
\DeclareUnicodeCharacter{"03BE}{\xi}
\DeclareUnicodeCharacter{"03C0}{\pi}
\DeclareUnicodeCharacter{"03C1}{\rho}
\DeclareUnicodeCharacter{"03C3}{\sigma} 
\DeclareUnicodeCharacter{"03C4}{\tau}
\DeclareUnicodeCharacter{"03C5}{\upsilon}
\DeclareUnicodeCharacter{"03C6}{\phi}
\DeclareUnicodeCharacter{"03D5}{\varphi}
\DeclareUnicodeCharacter{"03C7}{\chi}
\DeclareUnicodeCharacter{"03C8}{\psi}
\DeclareUnicodeCharacter{"03C9}{\omega}
\DeclareUnicodeCharacter{"21D0}{\Leftarrow}
\DeclareUnicodeCharacter{"0212B}{\AA}
\DeclareUnicodeCharacter{"00B7}{\cdot}
\DeclareUnicodeCharacter{"266A}{\eighthnote}
\DeclareUnicodeCharacter{"266B}{\twonotes}
\PrerenderUnicode{Σ}\PrerenderUnicode{×}
\PrerenderUnicode{̂}

\definecolor{title}{rgb}{0.3,0.5,0.9}
\definecolor{abst}{rgb}{0.366,0.366,0.266}
\definecolor{sect}{rgb}{0.2,0.5,0.8}
\definecolor{ssect}{rgb}{0.1,0.5,0.7}
\definecolor{sssect}{rgb}{0.0,0.5,0.6}
\definecolor{appsect}{rgb}{0.1,0.5,0.5}
\definecolor{ref}{rgb}{0.2,0.5,0.4}
\newcommand{\Title}[1] {\title{\color{title}\Huge #1}}
\newcommand{\TPheader}[3] {\date{}\maketitle\thispagestyle{fancy}\pagenumbering{alph}\lhead{#1}\chead{#2}\rhead{#3}\cfoot{}}
\newcommand{\makepage}[1] {\newpage\pagenumbering{#1}}

\newcommand{\Abstract}[1] {\begin{abstract}\normalsize #1 \end{abstract}}

\sectionfont{\color{sect}\Large\bf}
\subsectionfont{\color{ssect}\large\bf}
\subsubsectionfont{\color{sssect}\normalsize\bf}
\renewcommand{\appendix}{\setcounter{section}{0}\sectionfont{\color{appsect}\Large\bf}\renewcommand{\thesection}{\Alph{section}}\renewcommand*{\theHsection}{app.\the\value{section}}} 
\newcommand\references[1]{\sectionfont{\color{ref}\Large\bf}\bibliographystyle{hephys}\bibliography{#1}}

\newcommand\eqs[1] {\begin{align}#1\end{align}}

\newcommand\eqst[1] {\begin{multline}#1\end{multline}}

\newcommand\eqsc[1] {\begin{gather}#1\end{gather}}
\newcommand\eqscn[1] {\begin{gather*}#1\end{gather*}}
\newcommand\equ[1] {\begin{equation}#1\end{equation}}

\newcommand\fig[2] {\begin{figure}[#1]\centering #2\end{figure}}
\newcommand\bmat[1] {\begin{bmatrix}#1\end{bmatrix}}

\newcommand\tabl[2] {\begin{table}[#1]\centering #2\end{table}}

\renewcommand\exp[1] {e^{#1}}
\renewcommand\i {\dot{\iota}}
\newcommand\half {\tfrac{1}{2}}
\newcommand\s {\sigma}

\renewcommand\( {\left(}
\renewcommand\) {\right)}
\newcommand\wh {\widehat}

\DeclareMathOperator{\Vol}{Vol}

\newcommand\B {{\mathcal B}}

\newcommand\I {{\mathcal I}}

\newcommand\N {{\mathcal N}} 

\renewcommand\P {{\mathcal P}}
\newcommand\V {{\mathcal V}}
\newcommand\W {{\mathcal W}}

\newcommand\fg {{\mathfrak g}}
\newcommand\fm {{\mathfrak m}}
\newcommand\fn {{\mathfrak n}}

\newcommand\nn {\nonumber\\}

\newcommand\Vb {\bar{\V}}
\newcommand\Ib {\bar{\I}}
\newcommand\Ao {\texttt{A1}}
\newcommand\At {\texttt{A2}}

\numberwithin{equation}{section} 
\begin{document}
\Title{Twisted Indices of more 3d Quivers}

\author{Dharmesh Jain\footnote{\href{mailto:d.jain@saha.ac.in}{d.jain@saha.ac.in}}\bigskip\\
\emph{Theory Division, Saha Institute of Nuclear Physics}\\ \emph{1/AF Bidhan Nagar, Kolkata 700064, India}
}

\TPheader{}{\today}{}

\Abstract{We continue the study of 3d $\N=2$ Chern-Simons (CS) quiver gauge theories on $Σ_{\fg}×S^1$. Using localization results, we compute the twisted index of recently constructed SCFTs in the large rank limit. According to AdS/CFT correspondence, this field theory computation gives a prediction for two quantities corresponding to their holographic duals: the volumes of certain 7-dimensional Sasaki-Einstein manifolds and the entropy of black holes in AdS${}_4×Y_7$.
}

\tableofcontents
\makepage{arabic}

\section{Introduction}\label{sec:IO}
In this short note, we continue the large rank analysis of Chern-Simons (CS) quiver gauge theories with non-uniform ranks from \cite{Jain:2019lqb} and apply it to non-$\wh{ADE}$ quivers, discussed in \cite{Amariti:2019pky}. A distinguishing feature of most of these theories is the presence of varying number of adjoint matter multiplets at each node of the quiver. We will compute the twisted index of such theories with generic chemical potentials $ν$'s (and flavour fluxes $\fn$'s) and verify that the expected results from \cite{Jain:2019lqb} continue to hold:
\begin{enumerate}
\item $F_{S^3}[Δ]=4\V[ν]$ or more practically, $\frac{1}{μ[Δ]^2}=\frac{1}{16\tilde{μ}[ν]^2}$ upon identifying $Δ=2ν$.\footnote{The $μ$ and $\tilde{μ}$ functions are Lagrange multipliers used to compute the $S^3$ free energy ($F_{S^3}$) and Bethe potential ($\V$), respectively. It turns out that $F_{S^3}\propto μ$ and $\V\propto\tilde{μ}$.} We do not explicitly write $F_{S^3}$ (depending on generic $R$-charges $Δ$'s) in this note but we have checked this explicitly for all the quivers discussed here.
\item $\I=(\fg-1)\frac{4πN^{\sfrac{3}{2}}}{3}\tilde{μ}^3\Big[\frac{4}{\tilde{μ}^2} -\frac{1}{2}∑_I'(\fn_I-2ν_I)\frac{∂(\tilde{μ}^{-2})}{∂ν_I}\Big]$.
\item The integral expression \eqref{indexpOfS} for $\I$ matches the $\I$ obtained directly from $\tilde{μ}$ as above, given the algorithm to extract $Y(x)$'s mentioned in \cite{Jain:2019lqb} is used.
\end{enumerate}

We expect that the extremization of these twisted indices with respect to $ν$'s reproduces the macroscopic entropy of the dual black hole solutions in the 4d gauged supergravity uplifted to M-theory. The explicit construction of these M-theory duals would be interesting to test the expressions given here. We leave this for future work.

\paragraph{Comment.} The authors of \cite{Amariti:2019pky} claim that the algorithm they prescribe (let's call it \At) to solve the relevant matrix models is universal in contrast to the one used in \cite{Gulotta:2011vp,Crichigno:2012sk,Jain:2019lqb} (let's refer to it as \Ao). One issue they point out is that the authors of {\Ao} \emph{``⋯just terminate the eigenvalue density at the point where all of the $δy_{(aI,bJ)}$ saturate ⋯ However we have observed that, ⋯ terminates only when the eigenvalue density becomes zero''}. This is misleading because the algorithm of \cite{Crichigno:2012sk} explicitly includes this statement: \emph{``This process is iterated until all $y_a$'s are related, ⋯ or until the eigenvalue distribution terminates, i.e., $ρ(x)$=0''}. It just so happens that the matrix models for $\N=3$ $\wh{ADE}$ quivers do not have those extra regions. The $\N=2$ $\wh{A}$ quivers do have the extra regions and have been successfully solved with {\Ao} by determining where $ρ(x)$ vanishes\cite{Jain:2015kzj} (the $\N=2$ $\wh{DE}$ quivers continue to behave as $\N=3$ quivers).

One crucial ingredient of {\At} is an additional step of extremization in every region with respect to $x^*$ (boundary $x$-values) and $μ$. However, this step is rather computation-intensive and all the examples give the same result using either algorithm as the authors of {\At} clarify. Moreover, it can be argued that this extra step is quite redundant if one realizes that $x^*$'s get related to $μ$ in every region and the extremization with respect to $μ$ just leads to normalizability of $ρ(x)$, which fixes $μ$. After that, using the (understated) result of \cite{Gulotta:2011si}: $F_{S^3}\propto μ$ (or $\V\propto\tilde{μ}$ in the context of this note), one gets the desired result directly.

We provide more examples to showcase that {\Ao} is as \emph{universal} as {\At}, while being \emph{simpler} to execute in explicit computations.\footnote{In fact, either of these algorithms can be summarized as a simple three-step recipe: 1. Extremize -- Solve equations of motion; 2. Saturate -- Find validity of the solution; 3. Iterate -- Repeat till necessary. The step 1 is the main ingredient of the recipe, which is quite hard to work with in \At. The step 2 is more or less the secret ingredient and the step 3 is appreciated only when it is no longer needed (just like this metaphor).} We use it to compute the twisted indices of (almost) linear quivers with generic CS levels $k$'s and parameters $\{ν, \fn\}$'s.

\paragraph{Outline.} In Section \ref{sec:RTI} we briefly review the computation of twisted index in the large $N$ limit following \cite{Jain:2019lqb}. In Section \ref{sec:TIC} we provide the twisted indices for various quiver theories as shown in Table \ref{tab:con}. Finally, we generalize the known results for $F_{S^3}$ of $\N=3$ $\wh{E}$ quivers to both the $F_{S^3}$ and $\I$ of the $\N=2$ $\wh{E}$ quivers in Appendix \ref{App:EQver}.

\section{Reviewing Twisted Index}\label{sec:RTI}
The topologically twisted index is the $Σ_{\fg}×S^1$ partition function with a topological twist along the Riemann surface of genus $\fg$, $Σ_{\fg}$ \cite{Benini:2015noa,Benini:2016hjo,Closset:2016arn,Closset:2017zgf,Closset:2018ghr}. The main result reads:
\eqst{Z_{Σ_{\fg}×S^1}=\frac{1}{|\W|}∑_{\fm_a}∮\bigg(∏_a∏_{\mathclap{\text{Cartan}}}du_a\bigg)\B^{\fg}\,\exp{2π∑_a k_au_a·\fm_a} ∏_a\bigg(∏_{α∈G}\(1-e^{2πα(u_a)}\)^{1-\fg}∏_{α>0}(-1)^{α(\fm_a)}\bigg) \\
×∏_{I}∏_{ρ∈R_I}\bigg(\frac{e^{πρ(u_I)+π\i ν_I}}{1-e^{2πρ(u_I)+2π\i ν_I}}\bigg)^{ρ_I(\fm)+(\fg-1)\(\fn_I+(Δ_I-1)\)}\,,
\label{ZTIgenexp}}
where $u=\i\(∫_{S^1}A+\i\s\)$ are the holonomies and $\fm=\frac{1}{2π}∫_{Σ_{\fg}}F$ are the magnetic fluxes corresponding to the gauge group, $ν=\(∫_{S^1}A^{bg}+\i\s^{bg}\)$ are the holonomies (or chemical potentials) and $\fn$ are the fluxes for the background vector multiplet coupled to flavour symmetry such that $\fn(\fg-1)$ is integer-quantized. The real part of $ν$ is defined modulo 1 so we choose $ν$ to satisfy $0<ν<1$.

In the large $N$ limit, we have to first evaluate the Bethe potential and then the index. We only give here the relevant formulas from \cite{Jain:2019lqb} with appropriate modifications. Considering $\N=2$ quiver theories with gauge group $⊗_aU(n_aN)$ along with (bi)fundamental and adjoint matter multiplets, we get the following constraint for the large $N$ matrix model to be local:
\eqsc{n_a\left[1-N^{\text{adj}}_a +2{\textstyle ∑_{i=1}^{N^{\text{adj}}_a}}ν^i_{(a)}\right] =∑_{b|(a,b)∈E}(1-ν_{(a,b)}-ν_{(b,a)})n_b\,.
\label{NUconditionTTI}}
The Bethe potential in large $N$ limit reads
\eqst{\V ≈N^{\sfrac{3}{2}}∫dx ρ(x)\Bigg[2πx ∑_{a,I}k_a y_{a,I}(x) -\frac{1}{24π^2}ρ(x)\Bigg\{∑_{(a,b)∈E}∑_{I,J}\Big[\arg\big(\exp{2π\i(y_{a,I}-y_{b,J}+ν_{(a,b)}-\sfrac{1}{2})}\big) \\
×\Big(π^2 -\arg\big(\exp{2π\i(y_{a,I}-y_{b,J}+ν_{(a,b)}-\sfrac{1}{2})}\big)^2\Big) +(ν_{(b,a)}\text{ term})\Big] +∑_{a,I,J}∑_{i=1}^{N^{\text{adj}}_a}\Big[\arg\big(\exp{2π\i(y_{a,I}-y_{a,J}+ν_{(a)}^i -\sfrac{1}{2})}\big) \\
×\Big(π^2 -\arg\big(\exp{2π\i(y_{a,I}-y_{a,J}+ν_{(a)}^i -\sfrac{1}{2})}\big)^2\Big)\Big]\Bigg\} +π|x|(n_F-ν_F) \Bigg] -2π\tilde{μ}N^{\sfrac{3}{2}}\bigg(∫dx\,ρ(x) -1\bigg).
\label{BetheV}}
Here, $ν_F=∑_a∑_{\{f^a\}}n_a\(ν_{f^a}+\bar{ν}_{f^a}\)$ and we need to set $∑_an_ak_a=0$, $f^a =\bar{f}^a$. On general grounds\cite{Gulotta:2011si}, extremizing $\V$ gives
\equ{\bar{\V}=\frac{4πN^{\sfrac{3}{2}}}{3}\tilde{μ}\,.
\label{genextV}}

It turns out that the large $N$ limit of $\V$ is not enough to compute the twisted index because $\V$ has no divergences at leading order whereas the original Bethe Ansatz equations (BAEs) display divergent behaviour. This behaviour follows due to bifundamental and adjoint contributions involving $v'(z)$ diverging at $z=0$. Introducing exponentially small corrections in the Bethe equation leads to:
\eqst{\B_a^I ≈ -N^{\sfrac{3}{2}}ρ(x)\Bigg[∑_{b|(a,b)∈E}∑_J\Big[δ_{(δy_{ab,IJ}(x)+ν_{(a,b)},0)} Y^+_{(a,I;b,J)}(x) - δ_{(δy_{ab,IJ}(x)-ν_{(b,a)},0)} Y^-_{(a,I;b,J)}(x) \Big] \\
+∑_{J,±}∑_{i=1}^{N^{\text{adj}}_a}±δ_{(δy_{a,IJ}(x)±ν_{(a)}^i,0)} Y^±_{(a,I;a,J)}(x)\Bigg]\,,
\label{expYs}}
where $δ_{(f(x),0)}$ is the Kronecker delta symbol that equals 1 when $f(x)=0$ and 0 otherwise. The above equation is used to extract the $Y(x)$ functions (while keeping track of the sign) from (naïve) equations of motion $\B_a^I$ evaluated at the saturation values of the $δy(x)$'s as denoted by the $δ_{(δy(x)±ν,0)}$.

Finally, the matrix model for twisted index $\I$ leads to the following constraint on the flavour fluxes:
\eqsc{2n_a\left[1-N^{\text{adj}}_a +{\textstyle ∑_{i=1}^{N^{\text{adj}}_a}}\fn^i_{(a)}\right] =∑_{b|(a,b)∈E}(2-\fn_{(a,b)}-\fn_{(b,a)})n_b\,,
\label{FFconditionTTI}}
such that the integral expression for $\I$ reads
\eqs{\I &=\log|Z_{Σ_{\fg}×S^1}| ≈ (\fg-1)N^{\sfrac{3}{2}}∫dx ρ(x)\Bigg[\frac{1}{4π}ρ(x)\bigg(∑_{a,I,J}\arg\big(\exp{2π\i (y_{a,I}(x) -y_{a,J}(x) -\sfrac{1}{2})}\big)^2 \nn
&\quad -∑_{(a,b)∈E}∑_{I,J}\Big[\fn_{(b,a)}\arg\big(\exp{2π\i(y_{a,I}(x) -y_{b,J}(x)-ν_{(b,a)})}\big)^2 \Big] -(\fn_{(a,b)}\text{ term}) \nn
&\quad -∑_{a,I,J}∑_{i=1}^{N^{\text{adj}}_a}\Big[\fn_{(a)}\arg\big(\exp{2π\i(y_{a,I}(x) -y_{a,J}(x)-ν_{(a)}^i)}\big)^2 \Big] \bigg) +∑_{(a,b)∈E}∑_{I,J} δ_{(δy_{ab,IJ}(x)±ν_{(·,·)},0)}\fn_{(·,·)}Y^±_{(a,I;b,J)}(x) \nn
&\quad +∑_{a,I,J}∑_{i=1}^{N^{\text{adj}}_a} δ_{(δy_{a,IJ}(x)±ν^i_{(a)},0)}\fn^i_{(a)}Y^±_{(a,I;a,J)}(x) +π|x|(2n_F -\fn _F)\Bigg],
\label{indexpOfS}}
where $\fn_F$ is defined similar to $ν_F$. The above expression is to be evaluated by substituting $\big\{ρ(x),y_{a,I}(x),Y^±_{(a,I;b,J)}(x)\big\}$ obtained from extremizing the Bethe potential.

The Bethe potential $\V$ and $\I$ can be related in the present context by a straightforward generalization of the proof in \cite{Jain:2019lqb}:
\equ{\Ib=(\fg-1)\bigg[4\Vb +{∑_I}'(\fn_I-2ν_I)\frac{∂\Vb}{∂ν_I}\bigg]\,,
\label{genind}}
where the index $I$ now runs over both the bifundamental and adjoint matter multiplets, and the $'$ denotes the sum is over an independent set of chemical potentials. As it turns out, the $\tilde{μ}$ function encodes all the relevant quantities so we rewrite the twisted index as follows:
\equ{\bar{\I}=(\fg-1)\frac{4πN^{\sfrac{3}{2}}}{3}\tilde{μ}^3\bigg[\frac{4}{\tilde{μ}^2} -\frac{1}{2}{∑_I}'\(\fn_I -2ν_I\)\frac{∂\big(\frac{1}{\tilde{μ}^2}\big)}{∂ν_I}\bigg]\,.
\label{indexpOnS}}
This relation allows a direct check with \eqref{indexpOfS}. We also recall from \cite{Jain:2019lqb} that the Bethe potential is related to the free energy on $S^3$ as follows:
\equ{4\bar{\V}[ν]=\bar{F}_{S^3}[2ν]=\frac{4πN^{\sfrac{3}{2}}}{3}μ[2ν]\quad\text{ with }\quad \frac{1}{8μ^2}=\frac{1}{128\tilde{μ}^2}=\frac{\Vol(Y_7)}{\Vol(S^7)}\,,
\label{volmu}}
where $\Vol(Y_7)$ is the volume of the 7-dimensional Sasaki-Einstein manifold $Y_7$ appearing in the AdS${}_4×Y_7$ M-theory dual. Thus, we obtain not just the AdS${}_4$ black hole entropy but one more interesting quantity -- $\Vol(Y_7)$ -- from $\tilde{μ}$ via the AdS/CFT correspondence.\\

\noindent Let us now turn to explicit computation of the twisted index.

\section{Computing Twisted Index}\label{sec:TIC}
We consider quiver theories with gauge group $⊗_aU(n_aN)$ coupled to various matter multiplets in bifundamental and adjoint representations. The bifundamental multiplets are denoted by the edges in the quiver diagrams and adjoint multiplets by self-loops. As a shorthand, let us label these theories with $n_a$'s (`comarks') and number of adjoint multiplets coupled to the quiver nodes, as in $L^{\{n_a\},\{N^{\text{adj}}_a\}}$.\footnote{The $L$ could stand for Laufer, linear or lazy (to label lucidly).} Thus, $L^{\{1,2\},\{1,1\}}$ denotes a two-node quiver with comarks $\{1,2\}$ and 1 adjoint multiplet at each node. Table \ref{tab:con} tabulates the theories to be considered in this note.
\tabl{!h}{\caption{Quiver theories considered in specific (sub)sections below. The comarks $n_a$'s are written inside the nodes and CS levels $k_a$'s are marked beside each node.}
\begin{tabular}{cccc}
\hline
(Sub)Section\vphantom{\Big)} & Quiver & Label & (Sub)Section in \cite{Amariti:2019pky} \\
\hline\hline
\ref{sec:L1211} & $\vphantom{\bigg[}\vcenter{\hbox{\includegraphics[scale=1.5]{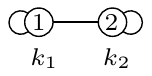}}}$ & $L^{\{1,2\},\{1,1\}}$ & 4.3 \\
\ref{sec:L1212} & $\vcenter{\hbox{\includegraphics[scale=1.5]{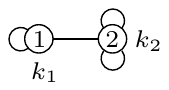}}}$ & $L^{\{1,2\},\{1,2\}}$ & 3, 4.2, 4.5 \\
\ref{sec:L112111} & $\vcenter{\hbox{\includegraphics[scale=1.5]{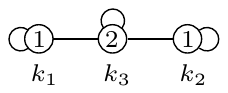}}}$ & $L^{\{1,1,2\},\{1,1,1\}}$ & 4.4 \\
\ref{sec:L11121110} & $\vcenter{\hbox{\includegraphics[scale=1.5]{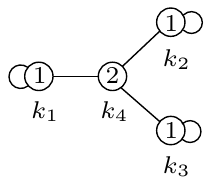}}}$ & $L^{\{1,1,1,2\},\{1,1,1,0\}}$ & 4.1 \\
\ref{sec:L112n} & $\vcenter{\hbox{\includegraphics[scale=1.5]{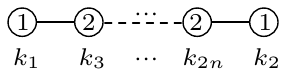}}}$ & $L^{\{1,1,2,⋯,2\},\{0\}}$ & A.5 \\
\ref{App:EQver} & See Figure \ref{fig:E678} & $\wh{E}_{6},\wh{E}_{7},\wh{E}_{8}\vphantom{\bigg]}$ & $-$ \\
\hline
\end{tabular}
\label{tab:con}}

\noindent We start this section with the discussion of the simplest quiver theory mentioned above.

\subsection[\texorpdfstring{$L^{\{1,2\},\{1,1\}}$}{L({1,2};{1,1})}]{$\bm{L^{\{1,2\},\{1,1\}}}$}\label{sec:L1211}
We choose $k_1=2k, k_2=-k$ without any loss of generality and the $ν$'s satisfy the following constraints coming from \eqref{NUconditionTTI}:
\equ{ν_{(1)}=1 -ν^+_{(1,2)}=4ν_{(2)}\qquad (0<ν_{(2)}<\tfrac{1}{4})\,.
}
We define $ν^±_{(a,b)}=ν_{(a,b)}±ν_{(b,a)}$. Similarly defining $\fn^±_{(a,b)}$, the $\fn$'s get constrained from \eqref{FFconditionTTI} as:
\equ{\fn_{(1)}=2 -\fn^+_{(1,2)}=4\fn_{(2)}\,.
}

Following the conventions and notations of \cite{Jain:2019lqb}, we present the solution of the matrix model for $L^{\{1,2\},\{1,1\}}$ below:
\paragraph{Region 1:} $\frac{-\mu(1 -4 ν_{(2)})}{k\(1 +ν^-_{(1,2)} -ν_{(2)}(5 -6ν_{(2)} +4ν^-_{(1,2)})\)} \leq x \leq \frac{\mu(1 -4 ν_{(2)})}{k\(1 -ν^-_{(1,2)} -ν_{(2)}(5 -6ν_{(2)} -4ν^-_{(1,2)})\)}$
\eqscn{\rho(x) = \frac{μ +k x ν^-_{(1,2)}}{2 ν_{(2)} \big(6 ν_{(2)}^2-5 ν_{(2)}+1\big)}
\,;\qquad y_{2,1}-y_{2,2}=0\,, \\
y_{1,1} - y_{2,2} = \frac{1}{2}\(\frac{k x \big(1 +ν_{(2)}(6ν_{(2)} -5)\big)}{μ +k x ν^-_{(1,2)}} -ν^-_{(1,2)}\).
}

\paragraph{Region $\bm{2^-}$:} $-\frac{\mu }{k (1 +ν^-_{(1,2)} -2ν_{(2)})} \leq x \leq \frac{-\mu(1 -4 ν_{(2)})}{k\(1 +ν^-_{(1,2)} -ν_{(2)}(5 -6ν_{(2)} +4ν^-_{(1,2)})\)}$
\eqscn{\rho (x) = \frac{μ +k x ν^-_{(1,2)}}{2 ν_{(2)} \big(6 ν_{(2)}^2-5 ν_{(2)}+1\big)}\,;\qquad y_{1,1}-y_{2,2}= -ν_{(1,2)}\,, \\
y_{2,1}-y_{2,2} = \frac{k x \(ν_{(2)}(5 -6ν_{(2)}) -1\)}{μ +k x ν^-_{(1,2)}} +4ν_{(2)} -1\,; \\
Y^+_{(1,1;2,2)} = -\frac{\pi \big(μ(1 -4ν_{(2)}) +k x (1 +ν^-_{(1,2)} -ν_{(2)}(5 -6ν_{(2)} +4ν^-_{(1,2)}))\big)}{ν_{(2)} \big(6 ν_{(2)}^2-5 ν_{(2)}+1\big)}\,\cdot 
}
\paragraph{Region $\bm{2^+}$:} $\frac{\mu(1 -4 ν_{(2)})}{k\(1 -ν^-_{(1,2)} -ν_{(2)}(5 -6ν_{(2)} -4ν^-_{(1,2)})\)} \leq x \leq \frac{\mu }{k(1 -ν^-_{(1,2)} -2ν_{(2)})}$
\eqscn{\rho(x) = \frac{μ +k x ν^-_{(1,2)}}{2 ν_{(2)} \big(6 ν_{(2)}^2-5 ν_{(2)}+1\big)}\,; \qquad y_{1,1}-y_{2,2}= ν_{(2,1)}\,, \\
y_{2,1}-y_{2,2} = \frac{k x \(ν_{(2)}(5 -6ν_{(2)}) -1\)}{μ +k x ν^-_{(1,2)}} -4ν_{(2)} +1\,; \\
Y^-_{(1,1;2,2)} = -\frac{\pi \big(μ(1 -4ν_{(2)}) -k x (1 -ν^-_{(1,2)} -ν_{(2)}(5 -6ν_{(2)} -4ν^-_{(1,2)}))\big)}{ν_{(2)} \big(6 ν_{(2)}^2-5 ν_{(2)}+1\big)}\,\cdot 
}

\paragraph{Region $\bm{3^-}$:} $-\frac{\mu }{k (1 +ν^-_{(1,2)} -3ν_{(2)})} \leq x \leq -\frac{\mu }{k (1 +ν^-_{(1,2)} -2ν_{(2)})}$
\eqscn{\rho (x) = \frac{μ +k x (1 +ν^-_{(1,2)} -3ν_{(2)})}{2 ν_{(2)}^2(1 -3ν_{(2)})}\,;\qquad y_{1,1}-y_{2,2}= -ν_{(1,2)}\,,\quad y_{2,1}-y_{2,2}= ν_{(2)}\,; \\
Y^+_{(1,1;2,2)} = -\frac{3 \pi  (1 -4 ν_{(2)}) (μ +k x (1 +ν^-_{(1,2)} -3ν_{(2)}))}{ν_{(2)} (1 -3 ν_{(2)})} -4 \pi  k x\,, \\
Y^-_{(2,1;2,2)} = -\frac{2 \pi  (μ +k x (1 +ν^-_{(1,2)} -2ν_{(2)}))}{ν_{(2)}}\,\cdot 
}
\paragraph{Region $\bm{3^+}$:} $\frac{\mu }{k(1 -ν^-_{(1,2)} -2ν_{(2)})} \leq x \leq \frac{\mu }{k(1 -ν^-_{(1,2)} -3ν_{(2)})}$
\eqscn{\rho (x) = \frac{μ -k x (1 -ν^-_{(1,2)} -3ν_{(2)})}{2 ν_{(2)}^2 (1 -3ν_{(2)})}\,;\qquad y_{1,1}-y_{2,2}= ν_{(2,1)}\,,\quad y_{2,1}-y_{2,2}= -\nu_{(2)}\,; \\
Y^-_{(1,1;2,2)} = -\frac{3 \pi  (1 -4 ν_{(2)}) (μ -k x (1 -ν^-_{(1,2)} -3ν_{(2)}))}{ν_{(2)} (1 -3 ν_{(2)})} +4 \pi  k x\,, \\
Y^+_{(2,1;2,2)} = -\frac{2 \pi  (μ -k x (1 -ν^-_{(1,2)} -2ν_{(2)}))}{ν_{(2)}}\,\cdot
}
The consistency of above solution demands $0\leq ν^-_{(1,2)}<\frac{1}{4}$ or $0<ν^-_{(1,2)} +3ν_{(2)}<1$. Integrating the eigenvalue density $ρ(x)$ then gives:
\equ{\frac{1}{\tilde{μ}^2}=\frac{2-5 \nu_{(2)}}{2k\nu_{(2)}(1 +\nu^-_{(1,2)} -2\nu_{(2)}) (1 -\nu^-_{(1,2)} -2\nu_{(2)}) (1 +\nu^-_{(1,2)} -3 \nu_{(2)}) (1 -\nu^-_{(1,2)} -3\nu_{(2)})}\,,
\label{ResLQ1}}
which leads to the expected relation of Bethe potential (given by \eqref{genextV}) and the $S^3$ free energy (to be computed similarly), $F_{S^3}=4\V$.\footnote{Specializing to the parameterization of \cite{Amariti:2019pky}, $ν_{(1,2)}=\half(1-ν)$, $ν_{(2)}=\frac{ν}{4}$, the above expression simplifies to the value mentioned there: $\frac{1}{\tilde{μ}^2}=\frac{32 (8-5ν)}{k ν (4-3ν)^2 (2-ν)^2}=\frac{16}{μ^2}\,·$} The above expression can be written in two familiar forms as follows:
\eqs{\frac{1}{\tilde{μ}^2} &=∑_{±} \frac{∓k}{\tilde{\s}_1^±\tilde{\s}_2^±}\,,\quad \text{ with }\quad\left\{
\begin{array}{l}
\tilde{\s}_1^± = k(ν^-_{(1,2)})\big(1 -2ν_{(2)} ±ν^-_{(1,2)}\big) \\
\tilde{\s}_2^± = k(4ν_{(2)})\big(1 -3ν_{(2)} ±ν^-_{(1,2)}\big)\,.
\end{array}\right. \\
\text{or }\; \frac{1}{\tilde{μ}^2} &=∑_{±,a=1}^2 \frac{2N_a}{\s_a^±}\,,\quad \text{ with }\quad\left\{
\begin{array}{ll}
N_1 = \frac{-1}{2ν_{(2)}^2(1-2ν_{(2)})}\,, & \s_1^± = 4k\big(1 -2ν_{(2)} ±ν^-_{(1,2)}\big) \\
N_2 = \frac{1}{2ν_{(2)}^2(1-3ν_{(2)})}\,, & \s_2^± = 4k\big(1 -3ν_{(2)} ±ν^-_{(1,2)}\big)\,.
\end{array}\right.
}
Note that when $ν^-_{(1,2)}=0$, a divergence is implied by $\tilde{\s}^±_1$ but that is spurious as is obvious from \eqref{ResLQ1}. Both these forms suggest an underlying polygon formulation as is well-known from $\wh{AD}$ quivers. In fact, the second form is easily obtained from differences of consecutive $ρ(x)$-values and the corresponding boundary $x$-values, which suggests that $ρ(x)$ can be interpreted as the height of this polygon. Since the double denominator form is tricky to obtain for larger quivers, we will present only the pure partial fractions form for quivers with three nodes and higher.

Finally, the twisted index follows from the relation \eqref{indexpOnS} and we have explicitly checked that this matches with the expression obtained by the integral in \eqref{indexpOfS}.

\subsection[\texorpdfstring{$L^{\{1,2\},\{1,2\}}$}{L({1,2};{1,2})}]{$\bm{L^{\{1,2\},\{1,2\}}}$}\label{sec:L1212}
This is the UV completion of Laufer theory, which can be denoted by $L^{\{1,2\},\{0,2\}}$ \cite{Laufer:1981,Aspinwall:2010mw}. We again choose $k_1=2k, k_2=-k$ and the $ν$'s satisfy the constraints following from \eqref{NUconditionTTI}:
\equ{ν_{(1)}=1-ν^+_{(1,2)}\,;\qquad ν^1_{(2)}+ν^2_{(2)}=\half+\tfrac{ν_{(1)}}{4}\,.
}
Similarly, the $\fn$'s are constrained from \eqref{FFconditionTTI} as follows:
\equ{\fn_{(1)}=2-\fn^+_{(1,2)}\,;\qquad \fn^1_{(2)}+\fn^2_{(2)}=1+\tfrac{\fn_{(1)}}{4}\,·
}
Note that by setting $ν_{(1)}=\half$ and $\fn_{(1)}=1$, we get the results for Laufer theory.

The matrix model for this theory will again have 3 regions and the computation proceeds as in the preceding section. Since this theory is symmetric under exchange of the two adjoints at node 2, we can evaluate the twisted index by assuming $0<ν^1_{(2)}<ν^2_{(2)}<1$ without loss of generality. With these, we can obtain the $\tilde{μ}$ function
\begingroup
\allowdisplaybreaks
\eqs{\frac{1}{\tilde{μ}^2}&=∑_{±}\frac{∓8k\big(1 -\nu_{(1)} +\nu^1_{(2)}\big)}{\tilde{\s}_1^±\tilde{\s}_2^±}\,, \\
&\quad \text{ with }\quad \left\{ \begin{array}{l}
\tilde{\s}_1^± = k(ν^-_{(1,2)})\big(1 -ν_{(1)} +ν^1_{(2)} ±ν^-_{(1,2)}\big) \\
\tilde{\s}_2^± = k\big(C_ν ±8\nu_{(1)}(1 -\nu_{(1)} +\nu^1_{(2)})ν_{(1,2)}^-\big)\,.
\end{array}\right. \nn
\text{or }\; \frac{1}{\tilde{μ}^2}&=∑_{±,a=1}^2 \frac{2N_a}{\s_a^±}\,, \\
&\quad \text{ with }\quad \left\{ \begin{array}{l}
N_1 = -\frac{16}{ν_{(1)}(4 -8ν_{(1)} +5{ν_{(1)}}^2) -4ν^1_{(2)}(4 -4ν_{(1)} +3{ν_{(1)}}^2) +8(4 -ν_{(1)}){ν^1_{(2)}}^2\vphantom{\big]}} \\
N_2 = -\frac{8ν_{(1)}(1 -ν_{(1)} +ν^1_{(2)})^2 N_1}{C_ν} \\
\s_1^± = 4k\big(1 -ν_{(1)} +ν^1_{(2)} ±ν^-_{(1,2)}\big) \\
\s_2^± = 4k\Big(\frac{C_ν}{8ν_{(1)}(1 -ν_{(1)} +ν^1_{(2)})} ±\nu^-_{(1,2)}\Big)\,.
\end{array}\right. \nonumber
}
\endgroup
Here, $C_ν = (\nu_{(1)}-2) \big(\nu_{(1)} (3\nu_{(1)}-2) -4 (2 +\nu_{(1)} -4\nu^1_{(2)})\nu^1_{(2)}\big)$. This solution is consistent for $ν^1_{(2)}<ν_{(1)}$ and $0<ν^1_{(2)}+ν^2_{(2)}<1$ along with $\s^±_1≥\s^±_2$. We have checked that $F_{S^3}=4\V$ for generic $ν$'s and the twisted index now follows from \eqref{indexpOnS}, which matches the one obtained from \eqref{indexpOfS} as expected. \emph{(This fact holds for all the following examples and we are not going to repeat this statement for them.)} There are two different parameterizations of $R$-charges (recall $Δ=2ν$) discussed in \cite{Amariti:2019pky} based on different superpotentials for $L^{\{1,2\},\{1,2\}}$ theory and the above result reproduces both cases as can be easily verified by direct substitution.

Furthermore, we can set $ν_{(1)}=\half$ as mentioned above to get the results for Laufer theory ($L^{\{1,2\},\{0,2\}}$). Setting $ν^-_{(1,2)}=0$, we can write a simpler expression for the $\tilde{μ}$ function (as its dependence can be reinstated from the definitions of $\s^±$):
\equ{\frac{1}{\tilde{μ}^2}=\frac{256 \big(11 +152 ν^1_{(2)} -160 {ν^1_{(2)}}^2\big)}{9 k \big(1 +2 ν^1_{(2)}\big) \big(1 +40ν^1_{(2)} -64 {ν^1_{(2)}}^2\big)^2}\,,
}
which reduces to $\frac{6656}{441 k}$ upon fixing $ν^1_{(2)}=\frac{3}{8}$. This value is nothing but $128$ times $\frac{\Vol(Y_7)}{\Vol(S^7)}$ given in \cite{Amariti:2019pky} as expected from \eqref{volmu}.

\subsection[\texorpdfstring{$L^{\{1,1,2\},\{1,1,1\}}$}{L({1,1,2};{1,1,1})}]{$\bm{L^{\{1,1,2\},\{1,1,1\}}}$}\label{sec:L112111}
We choose $k_1=-k_2-2k_3$ and the $ν$'s satisfy the constraints following from \eqref{NUconditionTTI}:
\equ{ν_{(a)}=1-ν^+_{(a,3)}\;\text{ for }a=1,2\,;\qquad ν_{(1)}+ν_{(2)}=4ν_{(3)}\,.
}
Similarly, the $\fn$'s are constrained from \eqref{FFconditionTTI} as follows:
\equ{\fn_{(a)}=2-\fn^+_{(a,3)}\;\text{ for }a=1,2\,;\qquad \fn_{(1)}+\fn_{(2)}=4\fn_{(3)}\,.
}
The matrix model solution will span 4 regions and there are three consistent solutions with the following saturation structure:
\equ{y_{(1,1)}→y_{(3,2)}\,;\quad y_{(2,1)}→y_{(3,2)}\,;\quad y_{(3,1)}→y_{(3,2)}\,.
\label{SB331}}
We call this a `branch' and present one of the solutions in this branch:
\begingroup
\allowdisplaybreaks
\eqsc{\frac{1}{\tilde{μ}^2}=∑_{±,a=1}^3 \frac{2N_a}{\s_a^±}\,, \label{mut33} \\
\text{ with } \left\{ \begin{array}{l}
\s_1^± = \big[2k_2\big(4ν_{(3)} -2 ±(ν^-_{(1,3)} -ν^-_{(2,3)})\big) +4k_3\big(4ν_{(3)} -2 ±ν^-_{(1,3)}\big)\big] \\
\s_2^± = \bmat{2k_2\big(4ν_{(3)} -2 ±(ν^-_{(1,3)} -ν^-_{(2,3)})\big) \\ +4k_3\Big((ν_{(1)} ±ν^-_{(1,3)}) -2\frac{ν_{(3)}}{ν_{(1)}}(1+2ν_{(1)} -3ν_{(3)})\Big)}\vphantom{\Bigg]} \\
\s_3^± = \bmat{2k_2\(±(ν^-_{(1,3)} -ν^-_{(2,3)}) +4\frac{ν_{(3)}}{ν_{(1)}ν_{(2)}}\({ν_{(1)}}^2 -2ν_{(3)}(1+2ν_{(1)}) +6{ν_{(3)}}^2\)\) \\ +4k_3\((ν_{(1)}±ν^-_{(1,3)})-2\frac{ν_{(3)}}{ν_{(1)}}(1+2ν_{(1)}-3ν_{(3)})\)}\vphantom{\Bigg]}
\end{array}\right. \nn
\text{ and } \left\{ \begin{array}{l}
N_1 = \frac{1}{(2ν_{(3)}-1)\(ν_{(1)}(2+ν_{(1)})-2ν_{(3)}(1+4ν_{(1)}-3ν_{(3)})\)} \\
N_2 = \frac{-{ν_{(1)}}^2}{\(ν_{(1)}(2+ν_{(1)})-2ν_{(3)}(1+4ν_{(1)}-3ν_{(3)})\)\({ν_{(1)}}^2 -4ν_{(1)}ν_{(3)}(1+ν_{(1)}) +4{ν_{(3)}}^2(1+4ν_{(1)} -3ν_{(3)})\)}\vphantom{\bigg]} \\
N_3 = \frac{ν_{(1)}ν_{(2)}}{\({ν_{(1)}}^2 -2ν_{(3)}(1+2ν_{(1)} -3ν_{(3)})\)\({ν_{(1)}}^2 -4ν_{(1)}ν_{(3)}(1+ν_{(1)}) +4{ν_{(3)}}^2(1+4ν_{(1)} -3ν_{(3)})\)}\vphantom{\bigg]}\,·
\end{array}\right. \nonumber
}
\endgroup
The consistency of this solution requires $ν_{(2)}>ν_{(3)}$, $1<ν_{(1)}+ν_{(2)}<2$ and $1<ν_{(2)}+ν_{(3)}<2$ along with $\s^±_1≥\s^±_2≥\s^±_3$.

There is another set of solutions which has the saturation structure as follows:
\equ{y_{(1,1)}→y_{(3,1)}\,;\quad y_{(2,1)}→y_{(3,2)}\,;\quad y_{(3,1)}→y_{(3,2)}\,.
\label{SB332}}
Both the branches \eqref{SB331} and \eqref{SB332} feature in \cite{Amariti:2019pky}, which uses a particular $R$-charge assignment and 3 different CS level assignments to give three results. The first and third cases are captured by the solutions in the first branch above, whereas the second case (and the first too) is given by those in the second branch, after relabelling nodes $1↔2$.

It is interesting to note that for $\wh{AD}$ quivers, there is a (re)parameterization of CS levels that leads to a single expression for $\tilde{μ}$ that captures all the solutions in any branch. So even though the solutions look quite different in the $k$-basis for this linear quiver, we except there is some parameterization of CS levels that leads to a `branch-invariant' expression for $\tilde{μ}$. Unfortunately, we leave this as open a problem as it has been for $\wh{E}$ quivers for quite some time.

\subsection[\texorpdfstring{$L^{\{1,1,1,2\},\{1,1,1,0\}}$}{L({1,1,1,2};{1,1,1,0})}]{$\bm{L^{\{1,1,1,2\},\{1,1,1,0\}}}$}\label{sec:L11121110}
We choose $k_1=-k_2-k_3-2k_4$ and the $ν$'s are constrained from \eqref{NUconditionTTI} to satisfy:
\equ{ν_{(a)}=1-ν^+_{(a,4)}\;\text{ for }a=1,2,3\,;\qquad ν_{(1)}+ν_{(2)}+ν_{(3)}=2\,.
}
Similarly, the $\fn$'s are constrained from \eqref{FFconditionTTI} as follows:
\equ{\fn_{(a)}=2-\fn^+_{(a,4)}\;\text{ for }a=1,2,3\,;\qquad \fn_{(1)}+\fn_{(2)}+\fn_{(3)}=4\,.
}
This quiver has a symmetry between nodes 2 and 3 and we assume $0<ν_{(2)}<ν_{(3)}<1$ without loss of generality.\footnote{The node 1 enjoys superficially special treatment as $k_1$ is written in terms of other CS levels. Though, permuting the nodes and imposing $∑_an_ak_a$ carefully restores the $Z_3$ symmetry.} The matrix model solution spans 5 regions and we give one of the solutions for $\tilde{μ}$ here:
\begingroup
\allowdisplaybreaks
\eqsc{\frac{1}{\tilde{μ}^2}=∑_{±,a=1}^4 \frac{2N_a}{\s_a^±}\,, \\
\text{ with } \left\{ \begin{array}{l}
\s_1^± = 2k_2\big(ν_{(3)}±(ν^-_{(1,4)}-ν^-_{(2,4)})\big) +2k_3\big(ν_{(2)}±(ν^-_{(1,4)}-ν^-_{(3,4)})\big) +4k_4\big(ν_{(2)}±ν^-_{(1,4)}\big) \\
\s_2^± = 2k_2\big(ν_{(3)}±(ν^-_{(1,4)}-ν^-_{(2,4)})\big) +2k_3\big(ν_{(2)}±(ν^-_{(1,4)}-ν^-_{(3,4)})\big) +4k_4\big(ν_{(3)}±ν^-_{(1,4)}\big) \\
\s_3^± = 2k_2\big(ν_{(3)}±(ν^-_{(1,4)}-ν^-_{(2,4)})\big) +2k_3\big(ν_{(1)}+ν_{(3)}±(ν^-_{(1,4)}-ν^-_{(3,4)})\big) +4k_4\big(ν_{(3)}+ν^-_{(1,4)}\big) \\
\s_4^± = 2k_2\big(ν_{(1)}+ν_{(2)}±(ν^-_{(1,4)}-ν^-_{(2,4)})\big) +2k_3\big(ν_{(2)}±(ν^-_{(1,4)}-ν^-_{(3,4)})\big) +4k_4\big(ν_{(2)}+ν^-_{(1,4)}\big)
\end{array}\right. \nn
\text{ and } \left\{ \begin{array}{l}
N_1 = \frac{1}{ν_{(2)}(ν_{(2)}-ν_{(3)})(ν_{(3)} -1)} \\
N_2 = \frac{1}{ν_{(3)}(ν_{(2)}-ν_{(3)})(ν_{(2)} -1)} \\
N_3 = \frac{1}{ν_{(3)}(ν_{(2)}+ν_{(3)} -2)(ν_{(2)} -1)} \\
N_4 = \frac{1}{ν_{(2)}(ν_{(2)}+ν_{(3)} -2)(ν_{(3)} -1)}\,·
\end{array}\right. \nonumber
}
\endgroup
The consistency conditions for this solution are $0<ν_{(1)}+ν_{(3)}>1$ along with $\s^±_1≥\s^±_2≥\s^±_3≥\s^±_4$. The latter leads to four conditions including $k_3≤0$ and $k_4≤0$. The other two involve all three CS levels and the adjoint chemical potentials as one may verify. This is a solution in the branch with saturations progressing as follows:
\equ{y_{(1,1)}→y_{(4,2)}\,;\quad y_{(2,1)}→y_{(4,2)}\,;\quad y_{(3,1)}→y_{(4,2)}\,;\quad y_{(4,1)}→y_{(4,2)}\,.
}

There can be one more branch (up to permutations of the three nodes with comark 1) involving the following saturations:\footnote{Note that even though we write ``saturations'', it does not necessarily mean the inequalities following from $\arg()$ functions in $\V$ get violated in every region. It just means two (or more) $y$'s get related by a $x$-independent value, which may or may not be the saturation value expected from the inequalities. In the latter case, one has to set $ρ(x)=0$ to get the final region boundary, as follows from a careful reading of {\Ao} \cite{Crichigno:2012sk,Jain:2019lqb}.}
\equ{y_{(1,1)}→y_{(4,1)}\,;\quad y_{(2,1)}→y_{(4,2)}\,;\quad y_{(3,1)}→y_{(4,2)}\,;\quad y_{(4,1)}→y_{(4,2)}\,.
}
Again, both these branches feature in \cite{Amariti:2019pky}, which uses a particular $R$-charge assignment and 2 different CS level assignments. We find that the first case is reproduced by the former branch and the second one by the latter branch upon permuting the nodes cyclically.

\subsection[\texorpdfstring{$L^{\{1,1,2,⋯,2\},\{0\}}$}{L({1,1,2,⋯,2};\{0\})}]{$\bm{L^{\{1,1,2,⋯,2\},\{0\}}}$}\label{sec:L112n}
This family of quivers has no adjoints and we will work out twisted indices for two members of the family. In general, $k_1=-(k_2 +2k_3 +2k_4+⋯2k_{2n})$ with $n≥2$ and the constraints following from \eqref{NUconditionTTI} and \eqref{FFconditionTTI} are:
\eqsc{ν^+_{(1,3)}=\half=ν^+_{(2n,2)}\,,\quad ν^+_{(3,4)}=\tfrac{1}{4}=ν^+_{(2n-1,2n)}\,,\quad ν^+_{(a,a+1)}=\tfrac{3}{4}\,,\;\text{ with } a=4,⋯,2n-2\,.\\
\fn^+_{(1,3)}=1=\fn^+_{(2n,2)}\,,\quad \fn^+_{(3,4)}=\half=\fn^+_{(2n-1,2n)}\,,\quad \fn^+_{(a,a+1)}=\tfrac{3}{2}\,,\;\text{ with } a=4,⋯,2n-2\,.
}

\paragraph{$\bm{n=2,\,L^{\{1,1,2,2\},\{0\}}}$.} We start by setting $ν_{(a,b)}=ν_{(b,a)}$ to evaluate the Bethe potential and restrict ourselves to the branch
\equ{y_{(1,1)}→y_{(3,2)}\,;\quad y_{(3,2)}→y_{(4,2)}\,;\quad y_{(3,1)}→y_{(4,2)}\,;\quad y_{(2,1)}→y_{(4,2)}\,;\quad y_{(4,1)}→y_{(4,2)}\,.
}
This gives us
\equ{\frac{1}{\tilde{μ}^2}=\frac{256}{5 k_2+6 k_3+8 k_4} -\frac{256}{3 (5 k_2+6 k_3+10 k_4)} -\frac{256}{3 (7 k_2+6 k_3+8 k_4)} -\frac{256}{5 (5k_2+10 k_3+10 k_4)}\,·
}
The consistency of the above solution requires $k_2≤k_4≤0$ and $2k_3 +k_4≤k_2$. We can similarly set $\fn_{(a,b)}=\fn_{(b,a)}$ to get the twisted index and we recover the expected relation $\I=4\V$.

To get the general result, we rewrite the above expression as:
\equ{\frac{1}{\tilde{μ}^2}=∑_{a=1}^4\frac{2N_a}{∑_{i=2}^4 c_a^i k_i} + \text{(same 4 terms),}\quad\text{ where} \qquad 
\begin{array}{c|cccc}
\hline
a & N_a & c_a^2 & c_a^3 & c_a^4 \\
\hline\hline
1 & 32 & \frac{5}{2} & 3 & 4 \\
2 & -\frac{32}{3} & \frac{5}{2} & 3 & 5 \\
3 & -\frac{32}{3} & \frac{7}{2} & 3 & 4 \\
4 & -\frac{32}{5} & \frac{5}{2} & 5 & 5 \\
\hline
\end{array}
\label{LinQ2}}
Now, we shift the $c_a^i$'s as follows
\equ{c_a^i → c_a^i ± ∑_{\mathclap{\substack{(a,b)∈\\ \P(1→i)}}}(-1)^{\s} 2n_i ν^-_{(a,b)}\,,
\label{LinQ2cs}}
such that the first set of four terms in \eqref{LinQ2} gets the shift with $+$ sign and the second set gets the $-$ sign. Here, $n_i$ are the comarks, $\P(1→i)$ denotes the set of edges taken along the path from node $1$ to node $i$ on the quiver diagram, and the sign $(-1)^{\s}$ is $+$ if the ordering of the label on $ν^-$'s is along the path and $-$ if it is reversed.\footnote{The significance of the sign $(-1)^{\s}$ is that retracing the path does not add any more $ν^-$'s than required. If there are multiple unique paths, then there are as many $\s^±$'s. This checks out for the quiver diagram of ABJM theory, which has two unique paths between its two nodes. One will need a slightly more sophisticated notation than in \eqref{LinQ2cs} for quivers with multiple edges between any given pair of nodes.} For example, the first term in the sum above would read in full detail as:
\equ{\frac{32}{(\frac{5}{2}+2ν^-_{(1,3)}+2ν^-_{(3,4)}+2ν^-_{(4,2)})k_2 + (3+4ν^-_{(1,3)})k_3 +(4+4ν^-_{(1,3)}+4ν^-_{(3,4)})k_4}\,·
}
If we had chosen to write $ν^-_{(2,4)}$ instead, the coefficient of $k_2$ would have $-2ν^-_{(2,4)}$, as the ordering $(2,4)$ of nodes is against the path $1→3→4→2$ when traversing from node 1 to node 2 on the quiver diagram (read the CS labels carefully in Table \ref{tab:con}).\footnote{It is post priori obvious that the $\s$'s defined for previous examples are normalized to have $±2n_iν^-_{(a,b)}$.} Note that introducing generic $ν_{(a,b)}$'s does not change the consistency conditions on CS levels here.

\paragraph{$\bm{n=3,\,L^{\{1,1,2,2,2,2\},\{0\}}}$.} The expression for $\tilde{μ}$ of this linear quiver is an expected generalization of the $n=2$ case:
\equ{\frac{1}{\tilde{μ}^2} = ∑_{±,a=1}^{6}\frac{2N_a}{\s^±_a}\,,\quad\text{ with}\qquad \begin{array}{c|cccccc}
\hline
a & N_a & c_a^2 & c_a^3 & c_a^4 & c_a^5 & c_a^6 \\
\hline\hline
1 & 32 & \frac{9}{2} & 3 & 4 & 7 & 8 \\
2 & -\frac{32}{3} & \frac{9}{2} & 3 & 4 & 7 & 9 \\
3 & -\frac{32}{3} & \frac{11}{2} & 3 & 4 & 7 & 8 \\
4 & -\frac{32}{5} & \frac{9}{2} & 3 & 4 & 9 & 9 \\
5 & -\frac{32}{45} & \frac{9}{2} & 3 & 9 & 9 & 9 \\
6 & -\frac{32}{27} & \frac{9}{2} & 9 & 9 & 9 & 9 \\
\hline
\end{array}
\label{LinQ3}}
Here, $\s^±_{a}=∑_{i=2}^{6}\big(c_a^i±∑_{(a,b)∈\P(1→i)}(-1)^{\s}2n_iν^-_{(a,b)}\big)k_i\,$. We chose the branch generalizing the one for $n=2$ case and the consistency conditions on CS levels follow from $\s^±_1≥\s^±_2≥⋯≥\s^±_6$.

\paragraph{General $\bm{n}$.} We can conjecture a general expression for $\tilde{μ}$ of these linear quivers as
\equ{\frac{1}{\tilde{μ}^2} = ∑_{±,a=1}^{2n}\frac{2N_a}{\s^±_a}\,,
}
with $\s^±_{a}=∑_{i=2}^{2n}\big(c_a^i±∑_{(a,b)∈\P(1→i)}(-1)^{\s}2n_iν^-_{(a,b)}\big)k_i\,$. We do not have much to say about $N_a$'s or $c^i_a$'s in general but it should be fun to generalize the two tables presented above in \eqref{LinQ2} and \eqref{LinQ3}. It would also be interesting to find a branch-invariant generic expression for this family of linear quivers.

\section*{\centering Acknowledgements}
DJ thanks P. M. Crichigno and A. Ray for collaboration on previous works that led to this note. He also acknowledges that most of this work would have been impossible without \texttt{Mathematica v11.3}\cite{Mathematica}.

\newpage
\appendix
\section[\texorpdfstring{$\wh{E}$ quivers}{Ê quivers}]{$\bm{\wh{E}}$ quivers}\label{App:EQver}
\fig{!h}{\includegraphics[width=\textwidth]{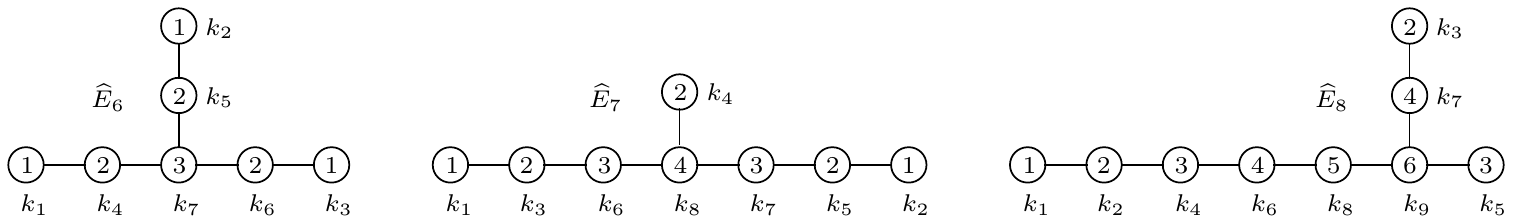}
\caption{$\wh{E}$ quivers with the comarks and CS levels marked.}
\label{fig:E678}
}
\noindent The quiver diagrams for these CS theories are shown in Figure \ref{fig:E678}. We will first focus on their $S^3$ free energy expressions given in \cite{Crichigno:2012sk}. First, note that the three formulas for $μ$ function of the $\N=3$ $\wh{E}_{n=6,7,8}$ can be written in the form:
\equ{\frac{1}{μ_{\wh{E}_n}^2}=\frac{2\(∑_{i=2}^{n+1}N^ik_i\)}{\(∑_{i=2}^{n+1}c_1^ik_i\)^2} -∑_{a=2}^n\frac{N_a}{\(∑_{i=2}^{n+1}c_a^ik_i\)} +\text{(same $n$ terms).}
\label{genEform}}
Of course, the tricky bit is to find the numerator factors $N_a$'s. If one knows the eigenvalue density $ρ(x)$, then it is easy to see that the $N_a$'s are basically the differences between $x$-independent pieces in $ρ(x)$ of adjacent regions (after setting $μ→1$). Let us rewrite the three $\N=3$ results with this knowledge:
\begingroup
\allowdisplaybreaks
\eqs{\frac{1}{\mu_{\wh{E}_6}^2}&=\frac{2 (4 k_2+11 k_3+8 k_4+4 k_5+6 k_6+4 k_7)}{(4 k_2+10 k_3+8 k_4+4 k_5+6 k_6+4 k_7)^2} \nn
&\quad -\frac{\frac{1}{6}}{(6 k_2+12 k_3+8 k_4+4 k_5+10 k_6+8 k_7)} \nn
&\quad -\frac{\frac{1}{42}}{(12 k_2+24 k_3+16 k_4+8 k_5+20 k_6+ 4 k_7)} \nn
&\quad -\frac{\frac{1}{154}}{(26 k_2+24 k_3+16 k_4+8 k_5+6 k_6+4 k_7)} \nn
&\quad -\frac{\frac{3}{22}}{\(4 k_2+\frac{28}{3} k_3+\frac{26}{3} k_4+8 k_5+6 k_6+4 k_7\)} \nn
&\quad -\frac{\frac{3}{2}}{\(4 k_2+\frac{28}{3} k_3+\frac{26}{3} k_4+4 k_5+6 k_6+4 k_7\)} +\text{(same 6 terms).}
}
\endgroup

\begingroup
\allowdisplaybreaks
\eqs{\frac{1}{\mu_{\wh{E}_7}^2}&=\frac{2(4 k_2+12 k_3+21 k_4+ 16 k_5+12 k_6+6 k_7+10 k_8)}{(4 k_2+12 k_3+20 k_4+16 k_5+12 k_6+6 k_7+10 k_8)^2} \nn
&\quad -\frac{\frac{1}{60}}{(24 k_2+36 k_3+48 k_4+36 k_5+24 k_6+12 k_7+24 k_8)} \nn
&\quad -\frac{\frac{1}{15}}{(4 k_2+11 k_3+18 k_4+16 k_5+14 k_6+12 k_7+9 k_8)} \nn
&\quad -\frac{\frac{4}{21}}{\(4 k_2+\frac{25}{2} k_3+21 k_4+16 k_5+11 k_6+6 k_7+\frac{27}{2} k_8\)} \nn
&\quad -\frac{\frac{9}{14}}{\(4 k_2+\frac{34}{3} k_3+\frac{56}{3} k_4+16 k_5+\frac{40}{3} k_6+6 k_7+10 k_8\)} \nn
&\quad -\frac{\half}{(4 k_2+12 k_3+20 k_4+18 k_5+12 k_6+6 k_7+10 k_8)}\nn
&\quad -\frac{\half}{(4 k_2+14 k_3+20 k_4+16 k_5+12 k_6+6 k_7+10 k_8)} +\text{(same 7 terms).}
}
\endgroup

\begingroup
\allowdisplaybreaks
\eqs{\frac{1}{\mu_{\wh{E}_8}^2}&=\frac{2(4 k_2+12 k_3+24 k_4+37 k_5+46 k_6+24 k_7+32 k_8+16 k_9)}{(4 k_2+12 k_3+24 k_4+36 k_5+46 k_6+24 k_7+32 k_8+16 k_9)^2} \nn
&\quad -\frac{\frac{1}{420}}{(60 k_2+90 k_3+120 k_4+150 k_5+180 k_6+90 k_7+120 k_8+60 k_9)} \nn
&\quad -\frac{\frac{3}{308}}{\(4 k_2+\frac{34}{3}(3 k_3+4 k_4+5 k_5+6 k_6+3 k_7+4 k_8+2 k_9)\)} \nn
&\quad -\frac{\frac{3}{55}}{\(4 k_2+12 k_3+\frac{70}{3} k_4+\frac{104}{3} k_5+46 k_6+23 k_7+\frac{103}{3} k_8+\frac{68}{3} k_9\)} \nn
&\quad -\frac{\frac{9}{35}}{\(4 k_2+12 k_3+\frac{70}{3} k_4+\frac{104}{3} k_5+46 k_6+23 k_7+\frac{103}{3} k_8+16 k_9\)} \nn
&\quad -\frac{\frac{9}{14}}{\(4 k_2+12 k_3+\frac{70}{3} k_4+\frac{104}{3} k_5+46 k_6+\frac{76}{3} k_7+32 k_8+16 k_9\)} \nn
&\quad -\frac{\half}{(4 k_2+12 k_3+24 k_4+36 k_5+48 k_6+24 k_7+32 k_8+16 k_9)} \nn
&\quad -\frac{\half}{(4 k_2+12 k_3+26 k_4+36 k_5+46 k_6+24 k_7+32 k_8+16 k_9)} +\text{(same 8 terms).}
}
\endgroup

Now we can give the transition rules (similar to subsection \ref{sec:L112n}) to write down the $μ$ function of $\N=2$ $\wh{E}$ quivers as follows:
\equ{N^i → N^i ±∑_{\mathclap{\substack{(a,b)∈\\ \P(1→i)}}}(-1)^{\s}n_iΔ^-_{(a,b)}\,;\qquad
c_a^i → c_a^i ±∑_{\mathclap{\substack{(a,b)∈\\ \P(1→i)}}}(-1)^{\s}n_iΔ^-_{(a,b)}\,,
}
with the two sets of $n$ terms in \eqref{genEform} corresponding to two $±$ signs, respectively. The twisted index then follows by constructing the $\tilde{μ}$ function along with $Δ→2ν$ as given by \eqref{volmu}:
\equ{\frac{1}{\tilde{μ}[ν]_{\wh{E}_n}^2}=\frac{16}{μ[2ν]_{\wh{E}_n}^2}\,·
}
We have verified that the $\tilde{μ}$ function for $\wh{E}_{6}$ and $\wh{E}_{7}$ quivers satisfies the above relation explicitly and leave such a verification for $\wh{E}_8$ quiver to interested readers. \smiley

\newpage
\references{bib3d} 

\end{document}